# Proposal of Signaling by Interference Control of Delayed-Choice Experimental Setup

Masanori Sato

*Honda Electronics Co., Ltd.,*
*20 Oyamazuka, Oiwa-cho, Toyohashi, Aichi 441-3193, Japan*

Tel: +81 532 41 2574
Fax: +81 532 41 2093
E-mail: msato@honda-el.co.jp

**Abstract**   We propose a new signaling system using the experimental setup of Wheeler's delayed-choice experiment previously carried out by T. Hellmuth, H. Walther, A. Zajonc, and W. Schleich [Phys. Rev. A, 35, (1987), 2532]. In the delayed-choice experiment, the experimental setup shows a wave property or a particle property at the time when the experimental conditions of the wave-particle duality of photons are chosen. Choice signals can be used as transmitting signals and the wave-particle duality of photons is used as receiving signals. For example, if we choose the wave property of a photon as a transmitting signal, we detect the interference of the wave at the detector that can be used as a receiving signal. Therefore, the experimental setup of the delayed-choice experiment can transmit information through interference.

1.INTRODUCTION

   Delayed-choice experiments were first proposed by Wheeler [1]. **Figure 1** shows a conceptual diagram of the experiment. A photon enters the interferometer via beam splitter 1 and is recombined by beam splitter 2. The two detectors show an interference signature. Wheeler pointed out that if the delayed choice of insertion of beam splitter 2 is chosen before a photon reaches the position of beam splitter 2, if beam splitter 2 is used, we detect the wave property of interference. If beam splitter 2 is not present, we detect the particle property.

   Delayed-choice experiments were experimentally examined by Hellmuth et al. [2]. **Figure 2** shows a schematic diagram of the experimental setup of the spatial-interference experiment with an optical switch (Pockels cell) [2]. A photon enters the interferometer via beam splitter 1 and is recombined by beam splitter 2. In this experiment, under the condition in which the optical switch is "on" (there are two paths X and Y), interference is detected at detectors X and Y. For example, photon paths can be arranged only detector X detects the photons. On the other hand, under the condition in which the optical switch is "off" (there is only path Y), photons can be detected with a 50 % probability at detectors X and Y.

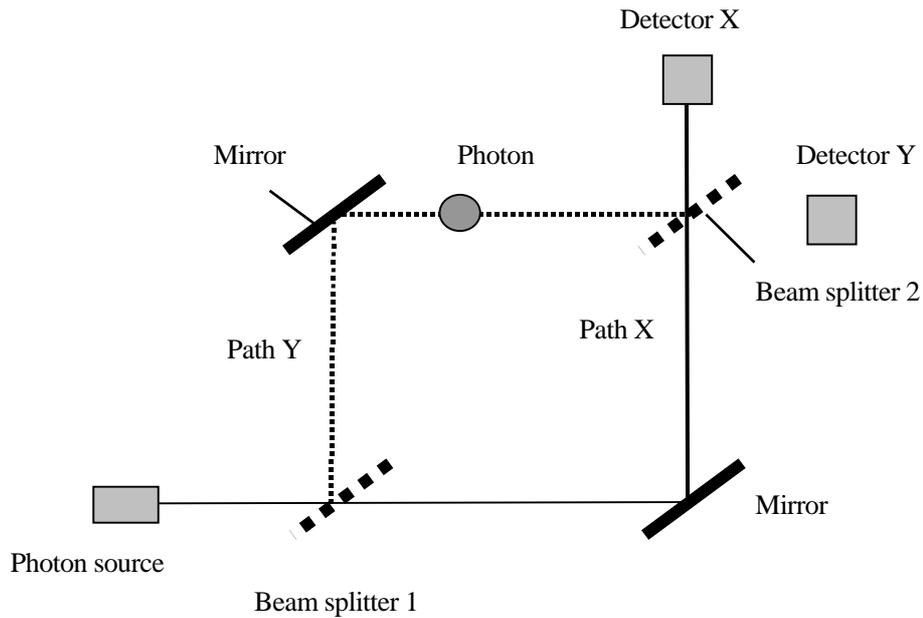

Fig. 1　Conceptual diagram of Wheeler's delayed choice experiments. Wheeler pointed out that when the delayed choice of insertion of beam splitter 2 is chosen before a photon reaches the position of beam splitter 2, if beam splitter 2 is used we detect wave property, if beam splitter 2 is not present we detect particle property.

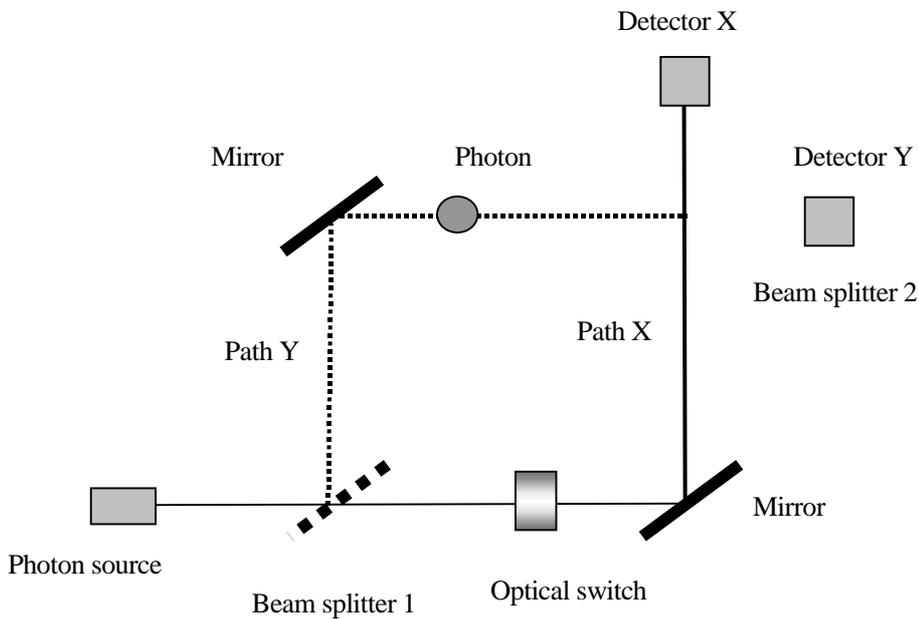

Fig. 2　Schematic diagram of experiment conducted by Hellmuth et al. [2] and proposal of signaling by delayed-choice experiment. We can choose the wave property under the condition of the optical switch is "on". We also choose the particle property under the condition in which the optical switch is "off". The optical switch works as a signal transmitter and detectors X and Y work as receivers. By choosing the property of wave-particle duality, we can transfer the information.

In the first proposal by Wheeler [1], there were no problems of causality. Delayed choice could be carried out at any time before a photon reached the position of beam splitter 2. Wave-particle duality was chosen at the time when a photon reached the position of beam splitter 2. If beam splitter 2 was used, wave property was detected, if beam splitter 2 was not present, particle property was detected.

However, in the experiments conducted by Hellmuth et al. [2] there arose the problem of causality. Delayed choice was performed using an optical switch, so there were photons that passed the optical switch and did not arrive at beam splitter 2. Some photons remained in the path between the optical switch and beam splitter 2. The behavior of these photons causes the problem of causality. There is a condition under which the optical switch decides the wave-particle duality of a photon remaining between the optical switch and beam splitter 2. We think that Hellmuth et al. [2] carefully examined the experimental conditions for satisfying the causality. They carefully eliminated the conditions that break the causality, so that the experimental data has complete compatibility with that of special relativity.

However, we consider that delayed-choice experiments show very interesting properties of wave-particle duality and causality. In this letter, we show that the delayed-choice experiment conducted by Hellmuth et al. [2] can be used to construct a signaling system, and discuss causality.

2. SIGNALING BY INTERFERENCE CONTROL
A. Experiment conducted by Hellmuth et al.

We show that the setup of the delayed-choice experiment conducted by Hellmuth et al. [2] can be used as a signaling system. In **Fig. 2**, we use the control signal of the optical switch as a transmitting signal: the "on" signal of the optical switch is 1, and the "off" signal is 0. We obtain information on the wave-particle duality of photons in the form of a receiving signal obtained at detectors X and Y: if we detect a signal only at detector X we know that the wave property is chosen, thus we obtain 1, and if the signal is detected with a 50 % probability at detectors X and Y (the particle property is chosen) we obtain 0. The person watching detectors X and Y can obtain information on whether the optical switch is "on" or "off". The signal that controls the optical switch can be used as a transmitting signal and we obtain information by watching detectors X and Y. We can use the optical switch as a transmitter and detectors X and Y as receivers. From the experimental data obtained by Hellmuth et al. [2], under light speed, we can transmit the information by interference control.

B. Experimental proposal of superluminal signaling

The experimental conditions that cannot satisfy causality were purposely eliminated in the experiment conducted by Hellmuth et al. [2] therefore we cannot obtain data that shows the possibility of superluminal signaling. Delayed choice is performed using the optical switch, so there are photons that pass the optical switch and do not arrive at beam splitter 2. There are photons that remain in the path between the optical switch and beam splitter 2. The behavior of these photons causes the problem of causality. The state of the optical switch decides the wave-particle duality of the photons between the optical switch and beam splitter 2. These experimental conditions should be tested experimentally.

At this stage, we consider that the interference depends on the state of the optical switch at the time a photon arrives at beam splitter 2. We think this conclusion is similar to that drawn in Wheeler's delayed-choice experiment. The delayed choice of optical switch operation should be performed before a photon reaches beam splitter 2. We obtain the results of the optical switch on: interference and the optical switch off: noninterference.

C. Transition time of Young's double slit pattern formation from Airy pattern

In Young's double slit experiment, when either slit of the double slits is closed, that is, when we carry out single slit experiment, Airy pattern is obtained. At the time when the closed slit is opened, Young's double slit pattern appears. Young's double slit pattern seems to be generated by transverse photon motion, which can be calculated from Bohm theory. That is, nonlocal quantum potential causes a small transverse displacement of photons. The transverse displacement of photons is enough smaller than the distance between the screen and the slit. Photon moves at the light speed, therefore the speed of the pattern formation looks to occur faster than light. That is, before the photon beam that passed the newly opened slit reaches the screen, Young's double slit pattern seems to be generated. If we use the slit condition (single or double) as transmitting signal and the interference pattern (Airy or Young) as receiving signal, signaling by interference control is possible. Signaling speed seems not to be restricted by the light speed. The displacement of photons is similar to group velocity and interference pattern formation is similar to phase velocity. We note that the experiment of Young's double slit pattern formation from Airy pattern should be carried out.

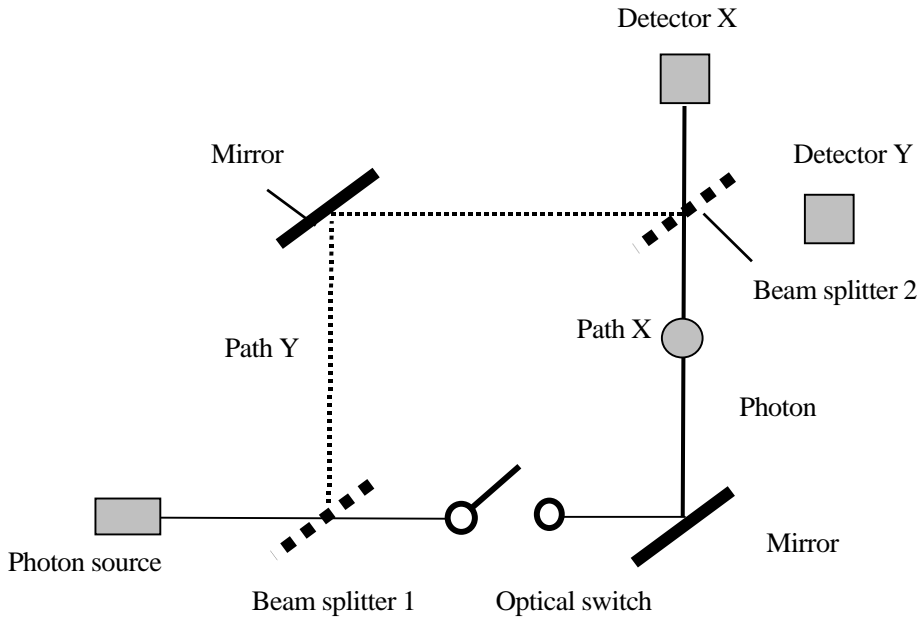

Fig. 3　Behavior of photon in path X after passing optical switch and before arriving at beam splitter 2. Interference at beam splitter 2 seems dependent on the state of the optical switch, that is, whether it is on or off, at the time when a photon reaches beam splitter 2.

## 3. DISCUSSION
### A. Causality

Hellmuth et al. [2] experimentally examined the delayed-choice experiment under the condition that causality was satisfied. They placed a 5 m glass fiber between beam splitter 1 and the optical switch. At the time a photon travels 1 m from beam splitter 1, the optical switch was controlled. The optical switch was controlled after the photon had passed beam splitter 1 and before photon arrival at the optical switch. Under these experimental conditions, the speed of information transmission was strictly restricted to be under light speed in the glass fiber.

**Figure 3** shows the case in which the optical switch is turned off when a photon is traveling between the optical switch and beam splitter 2. There were no experiments under this condition. At this stage, we believe Wheeler's argument that delayed choice performed before a photon reaches beam splitter 2 will show the result predicted from quantum mechanics. Therefore, a photon will not show interference at detectors X and Y. The person watching detectors X and Y knows that the optical switch is off. That is the information transmission. A photon travels at light speed in the glass fiber, however there is a problem of causality, i.e., information travels faster than light.

Of course, this conclusion has a problem of causality. There is a possibility that the state of the optical switch nonlocally controls the interference at beam splitter 2. Causality seems to be associated with a phenomenon in which a photon transfers information by itself, i.e., by the flight of the photon itself. However, during interference, the photon by itself does not transfer the information of interference. At this stage, we note that causality is a phenomenon that occurs through photons; information transmission that is transferred by the photon itself [3].

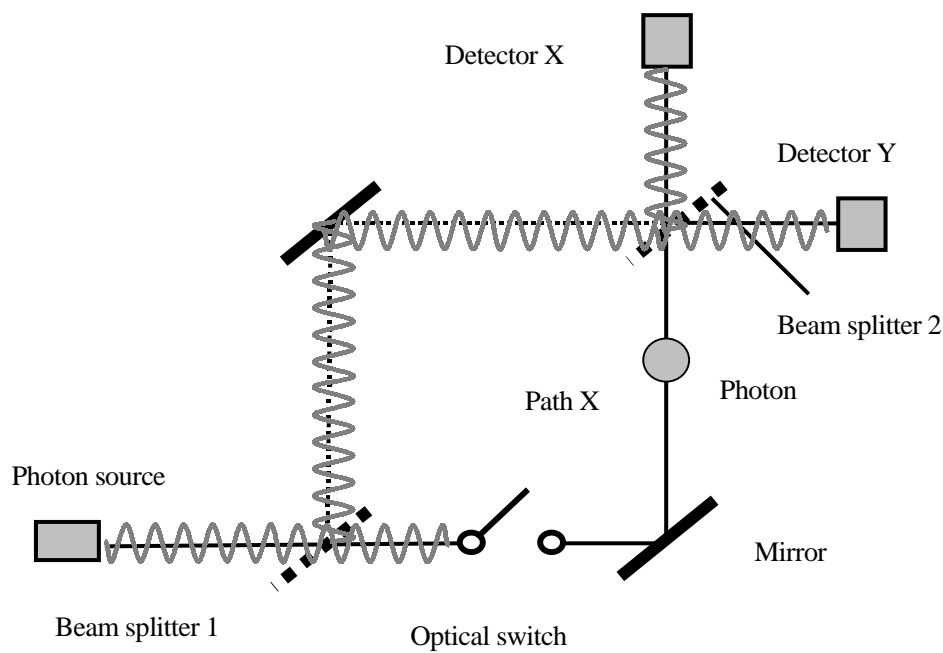

Fig. 4 Interpretation using de Broglie-Bohm picture; (a) The optical switch is off at the time a photon reaches beam splitter 2. There is no pilot wave, thus the particle property is expected.

B. De Broglie-Bohm picture

The de Broglie-Bohm picture [4, 5] gives a suitable interpretation of this situation. The phenomenon is explained by the wave and particle model. The wave is the pilot wave (quantum potential) and the particle is a photon that is guided by the pilot wave [4]. The wave is nonlocal and the particle is local. The photon travels at light speed, i.e., the photon satisfies causality; however, the wave does not satisfy causality (i.e., nonlocal). In **Fig. 4** (a), when the optical switch is turned off, there is no pilot wave at the photon position, so interference does not occur at beam splitter 2, then detectors X and Y detect photons with a 50 % probability. If the optical switch is turned on there is the pilot wave at the photon position and beam splitter 2 as shown in **Fig. 4** (b), thus we can detect interference. This means that the interference depends on the state of the optical switch at the time the photon arrives at beam splitter 2.

These experimental conditions should be tested experimentally. At this stage, we consider that the interference depends on the state of the optical switch at the time a photon arrives at beam splitter 2. As mentioned previously, we think this conclusion is similar to that drawn in Wheeler's delayed-choice experiment. The delayed choice of the optical switch operation should be performed before a photon reaches beam splitter 2. We obtain the results of the optical switch on: interference and optical switch off: noninterference. This conclusion seems to incorporate the problem of causality, however we consider that information transmission through the interference of photons can be eliminated from causality by restriction discussed in section A.

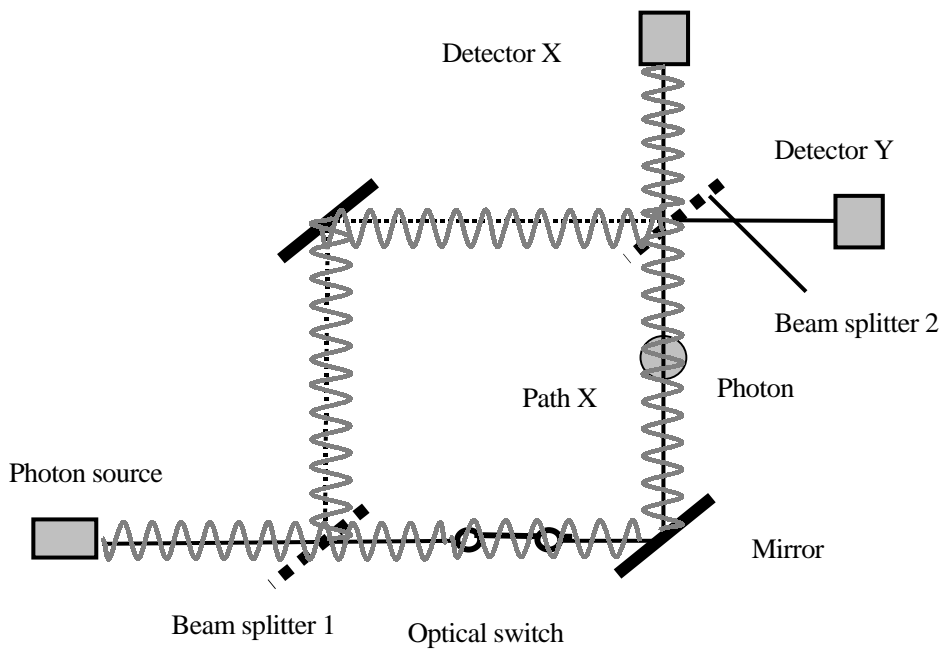

. Fig. 4   (b) The optical switch is on at the time a photon reaches beam splitter 2. There is a pilot wave, thus the wave property is expected.

## 4. CONCLUSIONS

We showed a method of signaling by interference control using the experimental setup of the delayed-choice experiment. In this experiment, information transmission is performed by the interference of photons. There is a very interesting problem of causality. In this experimental setup, causality is not well examined experimentally, as we have only considered a few previous experimental results. We do not have sufficient data to discuss the speed of information transmission. We consider that an experiment under the condition that a photon remains between the optical switch and beam splitter 2 should be carried out. At this stage, we consider the results that the interference depends on the state of the optical switch at the time the photon reaches beam splitter 2. Therefore not only signaling below light speed but also superluminal signaling can be performed using the delayed-choice experimental setup.